\documentclass[mypaper,8pt,twoside]{CoAst}
\usepackage{epsf,graphicx,fancyhdr}
\renewcommand{\chaptermark}[1]{\markboth{#1}{}}

\newcommand{\kopf}{\small\itshape Comm. in Asteroseismology \\ Contribution to the Proceedings of the Wroclaw HELAS Workshop, 2008}

\newcommand{\Authors}[1]{\begin{center}\normalsize\bf\sf #1 \end{center}}

\renewcommand{\author}[1]{\begin{center}\normalsize\bf\sf #1 \end{center}}
\newcommand{\Address}[1]{\begin{center}\small\sf #1 \end{center}}

\newcommand{\Session}[1]{{\vspace{3mm}\small \noindent  \hspace*{3mm} Session: } #1 \normalsize}

\newcommand{\Objects}[1]{{\vspace{0mm}\small \noindent  \hspace*{3mm} Individual Objects: } \small #1 \normalsize}

	\newcommand{\poster}{\small Poster \newline}

\newcommand{\References}[1]{\begin{flushleft}{\large References\\}\vspace*{2mm}\small #1 \end{flushleft}}

\newcommand{\chapterCoAst}[2]{\chapter[\sf\normalsize #1\\ \footnotesize \hspace*{5mm}by #2 \sf\normalsize][]{#1\\}\rhead[\fancyplain{}{\sf\footnotesize \center{#1}}]{\fancyplain{}{\sffamily\thepage}}\lhead[\fancyplain{\kopf}{\sffamily\thepage}]{\fancyplain{\kopf}{\sf\footnotesize \center{#2}}}}

\renewcommand{\chaptername}{}

\newcommand{\acknowledgments}[1]{\vspace*{5mm}\noindent  \textbf{Acknowledgments.} #1}

\def\rfr{\smallskip\par\noindent
        \hangindent=7truemm
        \hangafter=1}

\begin{document}
\sf

\chapterCoAst{New multisite observations of $\delta$ Scuti stars V624 Tauri and HD 23194}
{L.\,Fox Machado, E.\,Michel, M.\,Chevreton, et al.} 
\Authors{L.\,Fox Machado$^{1}$, E.\,Michel$^{2}$, M.\,Chevreton$^2$,
M.\,Alvarez$^1$, Z.P.\,Li$^3$, J.A.\,Belmonte$^4$,
A.\,Fernandez$^2$, L.\,Parrao$^1$, M.\,Rabus$^4$, J.\,Lochard$^4$,
F.\,P\'erez Hern\'andez$^4$, J.H.\,Pe\~na$^5$ and S.\,Pau$^2$}
\Address{$^1$ Observatorio Astron\'omico Nacional, Instituto de
Astronom\'{\i}a, Universidad Nacional Aut\'onoma de M\'exico, A.P.
877 Ensenada, BC 22860, Mexico\\
$^2$ Observatoire de Paris, LESIA, UMR 8109, F-92195,  Meudon,
France\\
 $^3$ Beijing Observatory, Chinese Academy of Sciences,
Beijing, P.R. China\\
$^4$ Instituto de Astrof\'{\i}sica de Canarias, E-38205
 La Laguna,\\ Tenerife, Spain\\
 $^5$ Instituto de Astronom\'{\i}a, Universidad
Nacional Aut\'onoma de M\'exico, Ap. P. 70-264, Mexico, D.F. 04510,
Mexico}

\Session{ \poster }
\Objects{V624\,Tauri, HD\,23194, HD\,23246}

\section*{Introduction}

 The stars  V624 Tau ($=$ HD 23156, BD$+$23$^{o}$ 495) and HD 23194
 ($=$ V1187 Tau, BD$+$24$^{o}$ 540) belong to the Pleiades cluster.
 While the former was identified as a $\delta$ Scuti variable
 by Breger (1972), the latter was classified as a $\delta$ Scuti pulsator
  by  Koen et al. (1999).  The multiperiodic pulsational behaviour of both stars was
 established as a result of the STEPHI X campaign in 1999 (Fox
 Machado et al. 2002).  In that campaign 7 frequencies for V624 Tau and
2 frequencies for HD 23194 were unambiguously detected above 99\%
confidence level. A comparison between the oscillation frequencies
and the eigenfrequencies of rotating models of some $\delta$ Scuti
stars of the Pleiades cluster,  V624 Tau and HD 23194 among them,
was carried out by Fox Machado et al. (2006). As a result, few
solutions with associated ranges of stellar parameters for each star
were found suggesting the existence of only $p$ modes, low radial
order in all the stars.

\medskip
  In order to increase  the number of detected modes in each star,
  a new STEPHI multisite campaign on V624 Tau and HD
 23194 was carried out in 2006. Some preliminary results of these observations are
given in this paper.

\section*{Observations, data reduction and frequency analysis}

\begin{table*}\centering
  \caption{Observational properties of the stars observed in the STEPHI
  2006
campaign.}
  \begin{tabular}{lcccccc}
  \hline
  Star        & HD    &  ST  & $V$ &  $B-V$   & $V \sin i$ & $\beta$ \\
              &       &      &        &       & $(\mathrm{km\, s}^{-1})$  \\
  \hline
  V624 Tauri      & 23156 & A7V      &  8.22 & +0.25 &  70 & 2.823  \\
  V1187 Tauri & 23194 & A5V      &  8.05 & +0.20  &  20 & 2.881  \\
  Comparison    & 23246 & A8V      &  8.17 & +0.27  & 200 & 2.773  \\
  \hline
  \end{tabular}
  \label{tab:stars}
\end{table*}

Some observational properties of the target stars are given in Table
1. The observations were carried out over the period 2006 November
14--December 3. As has been done in previous STEPHI campaigns, we
observed from three sites well distributed in longitude around the
Earth: Observatorio Astron\'omico Nacional (operated by the UNAM) in
San Pedro M\'artir, Baja California, Mexico; Xing Long Station
(operated by the Beijing Observatory) in Beijing, China; and
Observatorio del Teide (operted by the IAC) in Tenerife, Spain.
Four-channel photometers with interferometic blue filters were used
at all observatories. 232 hours of useful data were obtained during
20 nights of observations from the three sites. The efficiency of
observations was 48\% of the cycle.  The data reduction was obtained
following a classical scheme of multichannel photometry and is
similar to that reported in previous STEPHI campaigns (e.g. see
Alvarez et al. 1998, Fox Machado et al. 2002). The frequency peaks
of the light curves of the target stars were obtained by performing
a non linear fit to the data. Table 2 lists the peaks detected in
each star with a confidence level above 99\%. As can be seen eight
and three frequencies were detected in V624 Tau and HD 23194,
respectively. A combined analysis
 of the 1999 and 2006 STEPHI campaigns will be
given in a forthcoming paper (Fox Machado et al. 2008).

\begin{table}\centering
\caption{Frequency peaks detected above a 99\% confidence level in
our target stars. The origin of $\varphi$ is at HJD 2454040.97. S/N
is the signal-to-noise ratio after the pre-whitening process.}
\begin{tabular}{lccccr}
\hline
Star & & {\large $\nu $}      & A      & $\varphi$ & S/N  \\
     & & ($\mu$Hz)            & (mmag) & (rad)      \\
\hline
V624~Tau& $\nu_{1\mathrm{a}}$ & 243.01  & 3.37 & $+$1.4 & 5.0  \\
        & $\nu_{2\mathrm{a}}$ & 408.96 & 5.75 & $-$0.1  & 15.6   \\
        & $\nu_{3\mathrm{a}}$ & 413.20  & 4.15 & $-$0.1  & 11.5 \\
        & $\nu_{4\mathrm{a}}$ & 416.21  & 1.43 & $-$1.5  & 3.9   \\
        & $\nu_{5\mathrm{a}}$ & 451.80  & 3.82 & $-$1.4  & 10.4   \\
        & $\nu_{6\mathrm{a}}$ & 489.22  & 4.38 & $+$2.7  & 12.7   \\
        & $\nu_{7\mathrm{a}}$ & 529.18  & 1.43& $-$1.5  & 4.2   \\
        & $\nu_{8\mathrm{a}}$ & 690.92  & 1.06 & $-$2.9  & 4.3   \\
\hline
HD 23194 & $\nu_{1\mathrm{b}}$ & 533.62 & 6.94 & $-$1.8  & 14.5  \\
         & $\nu_{2\mathrm{b}}$ & 574.85  & 6.26 & $-$1.8  & 14.7  \\
         & $\nu_{3\mathrm{b}}$ & 615.40  & 4.50 & $-$0.8  & 12.2  \\
\hline
\end{tabular}
\label{tab:frequencies}

\end{table}

\acknowledgments{This work has received financial support from the
French CNRS, the Spanish DGES (AYA2001-1571, ESP2001-4529-PE and
ESP2004-03855-C03-03), the Mexican CONACYT and UNAM under grants
PAPIIT IN110102 and IN108106, the Chinese National Natural Science
Foundation under grant number 10573023 and 10433010. Special thanks
are given to the technical staff and night assistant of the Teide,
San Pedro M\'artir and Xing-Long Observatories and the technical
service of the Meudon Observatory.}

\References{

\rfr Alvarez M., Hern\'andez M.M., Michel E., et al. 1998, A\&A,
340, 149

\rfr Breger M. 1972, ApJ, 176, 367

\rfr Koen C., Van Rooyen R., Van Wyk F., et al. 1999, MNRAS, 309,
1051

\rfr Fox Machado L., \'Alvarez M., Michel E., et al. 2002, A\&A,
382, 556

\rfr Fox Machado L., Michel E., Chevreton M., et al. 2008, in
preparation

\rfr Fox Machado L., P\'erez Hern\'andez F., Su\'arez J.C., et al.
2006, A\&A, 446, 611

}

\end{document}